\documentclass[english]{article}
\usepackage[T1]{fontenc}
\usepackage[latin9]{inputenc}
\usepackage{color}

\makeatletter
\newcommand{\lyxaddress}[1]{
\par {\raggedright #1
\vspace{1.4em}
\noindent\par}
}

\makeatother

\usepackage{babel}
\begin{document}

\title{\textbf{GRAVITY WITH TORSION IN NON-CONSERVATIVE MAXWELL-LIKE GAUGE
APPROACH}}

\author{\textbf{G. Resconi$^{1*}$, I. Licata$^{2,3**}$, C. Corda $^{4***}$}}
\maketitle

\lyxaddress{\begin{center}
$^{1}$Catholic University, via Trieste 17 Brescia , Italy
\par\end{center}}

\lyxaddress{\begin{center}
$^{2}$ISEM, Inst. For Scientific Methodology, PA, Italy \& $^{3}$School
of Advanced International Studies on Applied Theoretical and Non Linear
Methodologies in Physics, Bari (Italy) 
\par\end{center}}

\lyxaddress{\begin{center}
\textbf{$^{4}$}Research Institute for Astronomy and Astrophysics
of Maragha (RIAAM), P.O. Box 55134-441, Maragha, Iran
\par\end{center}}

\begin{center}
\textit{E-mail addresses:} \textcolor{blue}{$^{*}resconi@numerica.it,\:{}^{**}ignazio.licata3@gmail.com$,
$^{***}cordac.galilei@gmail.com$} 
\par\end{center}
\begin{abstract}
In this work we take into consideration a generalization of Gauge
Theories based on the analysis of the structural characteristics of
Maxwell theory, which can be considered as the prototype of such kind
of theories (Maxwell-like). Such class of theories is based on few
principles related to different orders of commutators between covariant
derivatives. Their physical meaning is very simple, and lies in stating
that the local transformations of a suitable substratum (the space-time
or a particular phase space) and the imposed constraints define a
\textquotedblleft compensative mechanism\textquotedblright{} or the
\textquotedblleft interaction\textquotedblright{} we want to characterize.
After a mathematical introduction, we apply this approach to a modified
theory of gravity, in which the algebra of operators of covariant
derivatives leads to an additional term in the equation of motion
associated with the non-conservation of the energy-moment tensor.
This offers the possibility to include, without ad hoc physical assumptions
and directly from the formalism, new forms of coupling between matter
and energy and the expression of the mixing between gravity and torsion.
\end{abstract}
\begin{quotation}
\end{quotation}

\section{Introduction}

Symmetry and its physical implications on conservation principles
have a long history in Physics that goes back to the majestic work
by Emmy Nother \cite{key-1}. The same importance, if not higher,
is showed by the concepts of symmetry breaking and local gauge as
the constructive principle to characterize interactions as a \textquotedblleft compensation
mechanism\textquotedblright . In particular, all that made possible
a unified geometrical vision of fundamental interactions \cite{key-2,key-3}.
It is in such context, at the crossroad of Theoretical Physics, Cybernetics,
Category and Group Theory and Logical System Theory that the constructive
approach here introduced has been developed in \cite{key-4,key-5},
see also the recent\cite{key-6}. We stress the mathematical aspects
which make this approach a \textquotedblleft theory to build geometric-based
unified Theories\textquotedblright . The conceptual core of the procedure
can be expressed in a five point nutshell: 

a)The description of a suitable substratum and its global and local
properties on invariance; 

b)The field potentials are compensative fields defined by a gauge
covariant derivative. They share the global invariance properties
with the substratum; 

c)The calculation of the commutators of the covariant derivatives
in (b) provides the relations between the field strength and the field
potentials; 

d)The Jacobi identity applied to commutators provides the dynamic
equations satisfied by the field strength and the field potentials; 

e) The commutator between the covariant derivatives (b) and the commutator
(c) (triple Jacobian commutator) fixes the relations between field
strength and field currents. 

We re-analyse the modified theory of gravity in \cite{key-4}, in
which the algebra of the covariant derivatives operators adds a term
to the equations of motion like an eigenvalue equation of the double
commutator of covariant derivatives. We think such theory gets a strong
systemic value, because the GR syntax seems to regenerate itself from
inside and to produce many schemes of classical coupling Raum - Zeit
\textendash{} Materie. This auto-poietic feature is a distinctive
and propulsive of Theoretical Physics considered as a totality of
structures that fixes the conditions of thinkability for its entities
and \textquotedblleft beables\textquotedblright . We also underline
the conceptual meaning of gauging as cognitive compensation between
known and unknown domains \cite{key-7}. The proposed framework permits
also a novel approach to the controversial issue of torsion in gravity. 

\section{The Maxwell Dream }

As we know, Maxwell Theory has been the first model of gauge theory,
even if the awareness of its geometric nature came quite later of
its original formulation. In particular, Maxwell-like theories have
an easy physical interpretation as \textquotedblleft system theory\textquotedblright{}
, i.e. theories where the dynamic equations are those with the characterization
of their substratum. That is the case of General Relativity and -
at a more speculative level - of the M-Theory dreams, where each interaction
is a particular emergence of the world of brane {[}see for ex. 8{]}.
In principle, a criticism can be raised to this approach. Leading
historians such as Juergen Renn and Lee Smolin have suggested recently
that Einstein and the rest of us have learned the wrong lessons from
General Relativity because of the Einstein's retrospective misinterpretation
of his process of discovery. In particular, there exists a myth about
\textquotedblleft mathematical beauty\textquotedblright{} that should
be scaled down {[}9, 10{]}. Also such authors as Ogievetsky and Polubarinov,
decades ago, and Michael Deutsch et al., more recently, have shown
that we can have the same kind of massive Yang-Mills Theory justifying
the gauge approach by using \textquotedblleft nuts-and-bolts\textquotedblright{}
detailed reasoning from not-negotiable principles {[}for some recent
contributes see for ex. 11, 12, 13{]}.

A methodological rethinking about the ``non-negotiable'' principles
and the \textquotedblleft nuts-and-bolts\textquotedblright{} reasonings
is what we need in order to understand what we call here the \textquotedblleft Maxwellian
Dream\textquotedblright{} {[}14{]}. The Lagrangian approach is the
natural and historical extension of Newtonian Physics, based on the
concept of particle and its conservation principles, and the unequaled
model of \textquotedblleft nuts-and-bolts\textquotedblright{} physics.
With the advent of Quantum Mechanics (QM) and Quantum Field Theory
(QFT) in particular - as well as the strong conceptual problems Unified
Theories posed! - the Lagrangian-Newtonian scheme showed all its limitations.
First of all, the \textquotedblleft pointlike\textquotedblright{}
particle concept itself showed to be extremely complicated, just like
the attempts to define a particle \textquotedblleft structure\textquotedblright{}
{[}15, 16{]}. There are many recent experiments that seem to prove
against the existence of permanent property-bearing objects {[}17{]}.
We have to forget that the pointlike particle is at the root of the
Re-normalization method, born to eliminate some embarrassing infinitives
before becoming the key tool of the Effective Field Theories {[}18{]}.
Actually, also the non-Abelian gauge theories end up coinciding with
Super-Maxwell, thus the general schemes do not change so much. In
our opinion, the greatly elegant work by Deutsch does not imply that
the machineries of the standard model are better than the generalized
gauge principles. Instead, it shows the possibility to preserve the
global coherence of physics without considering problems about the
particle's structure. These problems are even stronger in the framework
of quantum gravity {[}19, 20{]}. Thus, as the general tapestry is
very indented and delicate, it is important to be careful for various
reasons, which are not merely aesthetic. Hence, we think that it is
interesting to explore also a generalized gauge approach, but without
considering the framework of the non-linear generalizations of the
QFT and without descriptions of the particle's structure in the sense
of \textquotedblleft collective emergency\textquotedblright . For
details on these issues, one can read the previously cited and precious
review of Pessa {[}14{]}. For example, a \textquotedblleft generalized
gauge\textquotedblright{} approach appears as the proper one to explore
unified theories by an \textquotedblleft offsetting\textquotedblright{}
method.

Historically, the Maxwell field seems to be the most \textquotedblleft trustwhorty\textquotedblright .
We could also take into consideration the \textquotedblleft hydrodynamic\textquotedblright{}
field, but it is less \textquotedblleft genuine\textquotedblright{}
and difficult to handle. We know that we pass from Classical Mechanics
to Maxwell one by means of a potential vector which opens Classical
Mechanics to something new. Such \textquotedblleft something\textquotedblright{}
manifests itself as a characteristic non-commutativity, a sort of
logical scheme \textquotedblleft openness\textquotedblright . In the
case of the potential vector, there comes out a new kind of derivative.
The non-commutativity of this new operator means that the new mechanic
systems cannot be integrated and that invariants are removed. Then,
one must create a gauge field (the electromagnetic field) which balances
out the simple mechanic phenomena and introduces the new field having
new invariants and new equations. Concluding, on one hand, the suggested
method is not founded on the creation of new Lagrangians. On the other
hand, it not the opposite. In fact, the structure of the Lagrangias
is derived through a process of subsequent openings and compensations.
The Einstein methods can be considered as a third \textquotedblleft middle
way''. In fact, on one hand one, one can use the Hilbert approach
in deriving the field equations, starting from the variation of the
action, which is a general and non-negotiable principle \cite{key-21}.
On the other hand, one cannot avoid the importance of the gauge principle
when operating in the linearized approximation or when discussing
the general covariance. In the case of the present paper, we stress
that the argument of the paper is the torsion in gravity, which is
currently a controversial and not completely understood issue. Thus,
we think that any approach, which could, in principle, help in a better
understanding of torsion in the framework of the gravitational theory
should not be avoided, because it is also useful in the framework
of the Alternative Theories of Gravity \cite{key-22}.

We know that the Maxwell equations in the tensor form are 

\begin{equation}
\begin{array}{c}
\partial_{\mu}F^{\mu\nu}=\frac{4\pi}{c}J^{\nu}\\
\\
\partial^{\alpha}F^{\mu\nu}+\partial^{\mu}F^{\nu\alpha}+\partial^{\nu}F^{\alpha\mu}=0,
\end{array}\label{eq: Maxwell equations}
\end{equation}
where the controvariant four-vector which combines electric current
density and electric charge density. $J^{\nu}=(cp,\,J_{x},\,J_{y},\,J_{z})$
is the four-current, the electromagnetic tensor is $F^{\mu\nu}=\partial^{\mu}A^{\nu}-\partial^{\nu}A^{\mu}$
with the four-potential $A^{\mu}=(\Phi,\,A_{x},\,A_{y},\,A_{z})$
containing the electric potential and vector potential. Using the
covariant derivative notation defined by 
\begin{equation}
D_{\mu}\equiv\partial_{\mu}-ieA_{\mu}\label{eq: covariant derivative}
\end{equation}
we have the classic relation 
\begin{equation}
\left[D_{\mu},\,D_{\nu}\right]=D_{\mu}D_{\nu}-D_{\nu}D_{\mu}=-ieF_{\mu\nu}.\label{eq: commutatore}
\end{equation}
Thus, we have 
\begin{equation}
F_{\mu\nu}=-F_{\nu\mu}.\label{eq: tensore}
\end{equation}
We know the tensor form of the Maxwell equations. They are invariant
for every affine transformation. In particular, the Maxwell equations
are invariant for Lorenz transformations. We remember that we can
obtain a general electromagnetic field from pure electrostatic field
with a suitable Lorenz transformation. For more general transformations,
the Maxwell equations are not invariant. Although rewriting the Maxwell
equations in a way to be invariant for any transformations of the
space time this is usually achieved by coupling Maxwell theory to
standard General Relativity, here we use a different approach. In
order to write the extension of the Maxwell equations, we define the
general commutators as in eq. (\ref{eq: commutatore}) where $F_{\mu\nu}$
is the general form for any field generated by the transformation.
Using eq. (\ref{eq: commutatore}), the equation 
\begin{equation}
\left[D_{\mu},\,\left[D_{\eta},\,D_{\nu}\right]\right]\psi+\left[D_{\eta},\,\left[D_{\nu},\,D_{\mu}\right]\right]\psi+\left[D_{\nu},\,\left[D_{\mu},\,D_{\eta}\right]\right]\psi=0\label{eq: commutatori}
\end{equation}
can be rewritten as
\begin{equation}
\left[D_{\mu},\,F_{\eta\nu}\right]\psi+\left[D_{\eta},\,F_{\nu\mu}\right]\psi+\left[D_{\nu},\,F_{\mu\eta}\right]\psi=0.\label{eq: commutatori 2}
\end{equation}
We have also the equation 
\begin{equation}
\left[D_{\gamma},\,\left[D_{\alpha},\,D_{\beta}\right]\right]\psi=\left[D_{\gamma},\,F_{\alpha\beta}\right]\psi=\chi J_{\gamma\alpha\beta}\psi\label{eq: J}
\end{equation}
In conclusion, the Maxwell scheme for a general gauge transformation
is 
\begin{equation}
\begin{array}{c}
\left[D_{\gamma},\,F_{\alpha\beta}\right]+\left[D_{\alpha},\,F_{\beta\gamma}\right]+\left[D_{\beta},\,F_{\gamma\alpha}\right]=0\\
\\
\left[D_{\gamma},\,F_{\alpha\beta}\right]\psi=\chi J_{\gamma\alpha\beta}\psi,
\end{array}\label{eq: Maxwell scheme}
\end{equation}
where $J_{\gamma\alpha\beta}$ are the currents of the particles that
generate the gauge field F. Such currents have the conservation rule
\begin{equation}
J_{\frac{\alpha\beta}{\gamma}}+J_{\frac{\beta\gamma}{\alpha}}+J_{\frac{\gamma\alpha}{\beta}}=0.\label{eq: conservation rule}
\end{equation}

\section{Maxwell-like Gauge Approach in Gravity}

The development of a Maxwellian program for gravity has been done
by Mignani-Pessa-Resconi and the recent developments about cosmological
constant have drawn the attention of an increasing number of researchers
to this theory \cite{key-4,key-5,key-6}. Some proposals concern also
application of the compensative method to the QM structure and to
system theory \cite{key-23,key-24}. The coupling of gravity with
a suitable substratum \textendash{} a minimal model of quantum physical
vacuum by a scalar field - implies the non conservation of the energy-momentum
tensor and allows a simple way to introduce the cosmological constant.
Let us label $\Omega$ the affine transformation from general coordinates
$x^{\mu}$ in a given frame to general coordinates $x^{\nu}$, that
is \cite{key-4}

\begin{equation}
x^{\nu}\rightarrow y^{\mu}=y^{\mu}(x^{\nu}).\label{eq: affine transformation}
\end{equation}
By differentiating eq. (\ref{eq: affine transformation}) one gets
the transformation law for controvariant vectors as \cite{key-4}
\begin{equation}
U^{\mu}=\frac{\partial y^{\mu}}{\partial x^{\nu}}V^{\nu}\label{eq: transformation law}
\end{equation}
together with the inverse ransformation 
\begin{equation}
V^{\nu}=\frac{\partial x^{\nu}}{\partial y^{\mu}}U^{\mu}.\label{eq: inverse}
\end{equation}
Considering the ordinary derivative $\partial_{\mu}=\frac{\partial}{\partial x^{\mu}}$,
the analysis in \cite{key-4} permits to write down the covariant
derivative as 
\begin{equation}
D_{\mu}\equiv\partial_{\mu}+\left[\frac{\partial x^{\nu}}{\partial y^{\mu}},\partial_{\mu}\right]\frac{\partial y^{\mu}}{\partial x^{\nu}},\label{eq: covariant derivative 2}
\end{equation}
which results to coincide with the ordinary covariant derivative of
General Relativity, see \cite{key-38} for a rigorours proof. Further,
in \cite{key-4} it has been shown that 
\begin{equation}
\left[D_{\mu},\,D_{\nu}\right]V_{\xi}=R_{\xi\nu\mu}^{\alpha}V_{\alpha}\label{eq: gravity commutator}
\end{equation}
where $R_{\delta\gamma\alpha}^{\beta}$ is the is the Riemann tensor.
One also obtains \cite{key-4} 
\begin{equation}
F_{\gamma\alpha}=\left[D_{\gamma},\,D_{\alpha}\right],\label{eq: commuta F}
\end{equation}
and inserts the gravitational matter current as \cite{key-4} 
\begin{equation}
J_{\gamma\alpha\beta}=\frac{\left[D_{\gamma},\,\left[D_{\alpha},\,D_{\beta}\right]\right]}{\chi}\label{eq: J 2}
\end{equation}
which is the gravitational analogous of eq. (\ref{eq: J}).

In this case, the Maxwell scheme gives us the set of equations \cite{key-4}
\begin{equation}
\begin{array}{c}
\left[D_{\gamma},\,R_{\alpha\beta}\right]+\left[D_{\alpha},\,R_{\beta\gamma}\right]+\left[D_{\beta},\,R_{\gamma\alpha}\right]=0\\
\\
\left[D_{\gamma},\,R_{\alpha\beta}\right]=\chi J_{\alpha\beta\gamma}
\end{array}\label{eq: Maxwell scheme for gravity}
\end{equation}
that together with eq. (\ref{eq: gravity commutator}) give \cite{key-4}
\begin{equation}
\begin{array}{c}
R_{\beta\gamma\delta}^{\alpha}+R_{\gamma\delta\beta}^{\alpha}+R_{\delta\beta\gamma}^{\alpha}=0\\
\\
D_{\epsilon}R_{\beta\gamma\delta}^{\alpha}+D_{\gamma}R_{\beta\delta\epsilon}^{\alpha}+D_{\delta}R_{\beta\gamma\epsilon}^{\alpha}=0\\
\\
\left[D_{\gamma},\,\left[D_{\alpha},\,D_{\beta}\right]\right]V_{\delta}=\left(D_{\gamma}R_{\delta\alpha\beta}^{\mu}\right)V_{\mu}+R_{\gamma\alpha\beta}^{\mu}D_{\mu}V_{\mu},
\end{array}\label{eq: Maxwell scheme for gravity 2}
\end{equation}
where the first equation is the first Bianchi identity, the second
is the second Bianchi identity and the last is the basic equation
for the Maxwell-like gauge approach in gravity \cite{key-4}. The
current can be given by the form 
\begin{equation}
\begin{array}{ccc}
J_{\alpha\beta\gamma}=D_{\alpha}T_{\beta\gamma}-D_{\beta}T_{\alpha\gamma}-\frac{1}{2}\left(g_{\beta\gamma}D_{\gamma}T-g_{\alpha\gamma}D_{\beta}T\right) & for\:which & D^{\gamma}J_{\alpha\beta\gamma}=0,\end{array}\label{eq:current}
\end{equation}
where $T_{\alpha\beta}$ is the energy-momentum tensor. The condition
$D_{\mu}V_{\mu}=0$ gives the gravitational field equation as \cite{key-4}
\begin{equation}
D^{\beta}R_{\alpha\beta}=\chi J_{\alpha\beta}^{\beta}\label{eq: gravitational field equation}
\end{equation}
Using the expression of the current one gets 
\begin{equation}
R_{\alpha\beta}=\chi\left(T_{\alpha\beta}-\frac{1}{2}g_{\alpha\beta}T\right)\label{eq: Einstein field equation}
\end{equation}
that is the Einstein field equation. 

\section{Torsion in non-conservative Gravity}

The most famous theory attempting to take into due account the presence
of torsion in gravity is the so called Einstein\textendash Cartan
theory, also known as the Einstein\textendash Cartan\textendash Sciama\textendash Kibble
theory \cite{key-26,key-27,key-28,key-29}. This is a (classical)
theory of gravitation similar to general relativity but with the important
difference that it assumes the presence of a torsion tensor, which
is the vanishing antisymmetric part of the the affine connection.
Thus, the torsion results coupled to the intrinsic spin of matter
in a similar way in which the curvature is coupled to the matter's
energy and momentum. This is because, in curved spacetime, the spin
of matter needs torsion to not be null, but to work as a variable
in the variational principle of stationary action. If one considers
the metric and torsion tensors as independent variables, one finds
the correct conservation law for the total (orbital plus spin) angular
momentum due to the presence of the gravitational field. Historically,
the theory was originally developed by Cartan, while additional contributions
are due to Sciama and Kibble. Einstein worked on this theory in 1928
through a failed approach in which torsion should match the electromagnetic
field tensor in the route for a unified field theory. At the end,
this approach led Einstein to the different theory of teleparallelism
\cite{key-30}. In general, the Einstein\textendash Cartan theory
is considered viable and there are various researchers who still works
on it, see for example \cite{key-31}. 

Torsion in gravity is currently considered a controversial issue.
For example, on one hand we find Hammond \cite{key-32}, who claims
that torsion is required for a complete theory of gravity, and that
without it, the equations of gravity violate fundamental laws. On
the other hand, Kleinert claims that torsion can be moved into the
curvature in a new kind of gauge transformation without changing the
physical content of Einstein\textquoteright s general relativity \cite{key-33}.
This should explain its invisibility in any gravitational experiment.
This controversy has an important scientific value, as the contenders
received the Fourth Award \cite{key-32} and a Honorable Mention \cite{key-33}
respectively in the important Gravity Research Foundation Essay Competitions
2010 and 2011 for their works on torsion in gravity. To introduce
the new wave equation for gravity we remember that in eq. (\ref{eq: gravity commutator})$D_{\mu}$
is the covariant derivative and the Riemann tensor is given by 
\begin{equation}
R_{\alpha\mu\nu}^{\lambda}\equiv\partial_{\mu}\Gamma_{\nu\alpha}^{\lambda}-\partial_{\nu}\Gamma_{\mu\alpha}^{\lambda}+\Gamma_{\mu\alpha}^{\sigma}\Gamma_{\nu\sigma}^{\lambda}-\Gamma_{\nu\alpha}^{\sigma}\Gamma_{\mu\sigma}^{\lambda}\label{eq: Riemann tensor}
\end{equation}
With the double commutator we have the third of eqs. (\ref{eq: Maxwell scheme for gravity 2})
as dynamic equation. Now, let us connect the commutator with the gravity
current as 
\begin{equation}
\chi J_{\mu\alpha\beta}=\left(D_{\mu}R_{\nu\alpha\beta}^{\lambda}\right)V_{\lambda}+R_{\mu\alpha\beta}^{\lambda}\left(\nabla_{\lambda}V_{\nu}\right),\label{eq: connect}
\end{equation}
When $\nabla_{\mu}V_{\nu}=0$ one can see through a bit of algebra
that the Einstein field equation (\ref{eq: Einstein field equation})
is retreived.

The original formulation of the general theory of relativity  is assumed
\textquotedbl{}torsion free\textquotedbl{}. On the other hand, based
on the increasing interest for the extended gravity theories, it is
useful to see how torsion appears in a modern geometrical  language.
We define the torsion tensor as 
\begin{equation}
T_{\mu\nu}^{\lambda}\equiv\Gamma_{\mu\nu}^{\lambda}-\Gamma_{\nu\mu}^{\lambda}\label{eq: Torsion}
\end{equation}
Now, we show in an explicit way that it is possible to present the
previous dynamical equation by a wave equation with particular source
where the variables are symmetric and anti-symmetric Christoffel symbols
including torsion in one geometric picture. We will define the new
type of wave with an explicit computation of the commutator and of
the double commutator. Thus, one gets 
\begin{equation}
D_{\mu}V_{\alpha}\equiv\frac{\partial V_{\alpha}}{\partial x_{\mu}}-\Gamma_{\mu\alpha}^{\lambda}V_{\lambda}.\label{eq: new type of wave}
\end{equation}
A bit of algebra permits to compute the first commutator as 
\begin{equation}
\begin{array}{c}
F_{\mu\nu,\alpha}=\left[D_{\mu},\,D_{\nu}\right]V_{\alpha}=D_{\mu}D_{\nu}V_{\alpha}-D_{\nu}D_{\mu}V_{\alpha}=\\
\\
-\left(R_{\alpha\mu\nu}^{\lambda}V_{\lambda}+T_{\mu\nu}^{\lambda}D_{\lambda}V_{\alpha}\right).
\end{array}\label{eq: first commutator}
\end{equation}
In conclusion one gets 
\begin{equation}
-F_{\mu\nu,\alpha}=\left[-D_{\mu},\,D_{\nu}\right]V_{\alpha}=G_{\mu\nu,\alpha}+\Omega_{\mu\nu,\alpha}\label{eq: first commutator 2}
\end{equation}
with 
\begin{equation}
\begin{array}{c}
G_{\mu\nu,\alpha}\equiv\left(\frac{\partial\Gamma_{\nu,\alpha}^{\lambda}}{\partial x_{\mu}}-\frac{\partial\Gamma_{\mu,\alpha}^{\lambda}}{\partial x_{\nu}}\right)V_{\lambda}\\
\\
\Omega_{\mu\nu,\alpha}\equiv\left(\Gamma_{\mu,\xi}^{\lambda}\Gamma_{\nu,\alpha}^{\xi}-\Gamma_{\mu,\alpha}^{\xi}\Gamma_{\nu,\xi}^{\lambda}\right)V_{\lambda}+\left(\Gamma_{\mu,\nu}^{\lambda}-\Gamma_{\nu,\mu}^{\lambda}\right)D_{\lambda}V_{\alpha}.
\end{array}\label{eq: imposizioni}
\end{equation}
Now, let us consider the rescaling 
\begin{equation}
\Gamma_{\mu,\nu}^{\lambda}\rightarrow\Gamma_{\mu,\nu}^{\lambda}+\frac{\partial\chi}{\partial x_{\mu}}.\label{eq: rescaling}
\end{equation}
One can easily show that $G_{\mu\nu,\alpha}$ is invariant under the
rescaling (\ref{eq: rescaling}). Now, we impose the Lorenz-like gauge
condition \cite{key-37} as 
\begin{equation}
\frac{\partial\Gamma_{\mu,\nu}^{\lambda}}{\partial x_{\mu}}=0.\label{eq: Lorenz condition}
\end{equation}
Some computations permits to obtain 
\begin{equation}
\frac{\partial G_{\mu\nu,\alpha}}{\partial x_{\beta}}+\frac{\partial G_{\beta\mu,\alpha}}{\partial x_{\nu}}+\frac{\partial G_{\nu\beta,\alpha}}{\partial x_{\mu}}=0.\label{eq: divergenza nulla}
\end{equation}
The Lagrangian gravitational density is 
\begin{equation}
\begin{array}{c}
L\equiv F_{\mu\nu,\alpha}F^{\mu\nu,\alpha}=G_{\mu\nu,\alpha}G^{\mu\nu,\alpha}+\\
\\
\Omega_{\mu\nu,\alpha}\Omega^{\mu\nu,\alpha}+2G_{\mu\nu,\alpha}\Omega^{\mu\nu,\alpha}
\end{array}\label{eq: Lagrangian gravitational density}
\end{equation}
where $G_{\mu\nu,\alpha}G^{\mu\nu,\alpha}$ and $\Omega_{\mu\nu,\alpha}\Omega^{\mu\nu,\alpha}+2G_{\mu\nu,\alpha}\Omega^{\mu\nu,\alpha}$
are the Lagrangian density for the free gravitational field and the
reaction field of the vacuum respectively. The interaction term connects
the gravitation field with the field of the vacuum. With a bit of
computing the dynamic equations for non conservative gravity can be
obtained as 
\begin{equation}
\begin{array}{c}
\left[D_{\beta},\,\left[D_{\mu},\,D_{\nu}\right]\right]V^{\alpha}=J_{\mu\nu,\alpha\beta}\\
\\
D_{\beta}G_{\mu\nu,\alpha}=-J_{\mu\nu,\alpha b}-D_{\beta}\Omega_{\mu\nu,\alpha}-\left[D_{\mu},\,D_{\nu}\right]D_{\beta}V_{\alpha}
\end{array}\label{eq. dynamic equations}
\end{equation}
with 
\begin{equation}
D_{\beta}G_{\mu\nu,\alpha}\equiv\frac{\partial G_{\mu\nu,\alpha}}{\partial x_{\beta}}-G_{j\nu,\alpha}\Gamma_{\mu\beta}^{\xi}-G_{\mu j,\alpha}\Gamma_{\nu\beta}^{\xi}-G_{\mu\nu,\xi}\Gamma_{\alpha\beta}^{\xi}.\label{eq: DG}
\end{equation}
Thus, 
\begin{equation}
\frac{\partial G_{\mu\nu,\alpha}}{\partial x_{\beta}}=-J_{\mu\nu,\alpha\beta}-D_{\beta}\Omega_{\mu\nu,\alpha}-\left[D_{\mu},\,D_{\nu}\right]D_{\beta}V_{\alpha}=J_{\mu\nu,\alpha\beta}.\label{eq: Thus}
\end{equation}
Eq. (\ref{eq: Thus}) implies 
\begin{equation}
\frac{\partial G_{\mu\nu,\alpha}}{\partial x_{\beta}}=\left(\frac{\partial^{2}\Gamma_{\nu,\alpha}^{\lambda}}{\partial x_{\beta}\partial x_{\mu}}-\frac{\partial^{2}\Gamma_{\mu,\alpha}^{\lambda}}{\partial x_{\beta}\partial x_{\nu}}\right)V_{\lambda}.\label{eq: implies}
\end{equation}
Setting $\partial x_{\beta=}\partial x_{\mu}$, one gets 
\begin{equation}
\frac{\partial G_{\mu\nu,\alpha}}{\partial x_{\mu}}=\left(\frac{\partial^{2}\Gamma_{\nu,\alpha}^{\lambda}}{\partial^{2}x_{\mu}}-\frac{\partial^{2}\Gamma_{\mu,\alpha}^{\lambda}}{\partial x_{\nu}\partial x_{\mu}}\right)V_{\lambda}.\label{eq: gets}
\end{equation}
The Lorentz-like gauge condition gives 
\begin{equation}
\frac{\partial^{2}\Gamma_{\mu,\alpha}^{\lambda}}{\partial x_{\nu}\partial x_{\mu}}=0.\label{eq: condizione}
\end{equation}
Then, 
\begin{equation}
\frac{\partial G_{\mu\nu,\alpha}}{\partial x_{\mu}}=\frac{\partial^{2}\Gamma_{\nu,\alpha}^{\lambda}}{\partial^{2}x_{\mu}}V_{\lambda}=\left(\frac{\partial^{2}\Gamma_{\nu,\alpha}^{\lambda}}{\partial^{2}x}-\frac{\partial^{2}\Gamma_{\nu,\alpha}^{\lambda}}{c^{2}\partial^{2}t}\right)V_{\lambda}=J_{\nu\alpha}.\label{eq: then}
\end{equation}
When the currents are equal to zero eq. (\ref{eq: then}) reduces
to 
\begin{equation}
\frac{\partial G_{\mu\nu,\alpha}}{\partial x_{\mu}}=\frac{\partial^{2}\Gamma_{\nu,\alpha}^{\lambda}}{\partial^{2}x_{\mu}}V_{\lambda}=\left(\frac{\partial^{2}\Gamma_{\nu,\alpha}^{\lambda}}{\partial^{2}x}-\frac{\partial^{2}\Gamma_{\nu,\alpha}^{\lambda}}{c^{2}\partial^{2}t}\right)V_{\lambda}=0.\label{eq: reduces}
\end{equation}
Hence, we find the interesting result that the derivative of the Christoffel
connection $\Gamma_{\nu,\alpha}^{\lambda}$ has wave behavior.

Now, some algebra computations permits to write the ``gravitational
currents'' as 
\begin{equation}
J_{\mu\nu,\alpha\beta}=\frac{1}{2}R_{\mu\nu,\alpha\beta}\;\Rightarrow J_{\mu\nu,\alpha\beta}=R_{\mu\nu,\alpha\beta}-J_{\mu\nu,\alpha\beta}\label{eq: =00201Cgravitational currents}
\end{equation}
where the second derivatives of the Riemann tensor represent a reaction
of a virtual matter or medium (vacuum) and the second derivatives
of $J$ are the ordinary currents for the ordinary matter represented
by the energetic tensor. The non-linear reaction of the self-coherent
system produces a current that justifies the complexity of the gravitational
field and non-linear properties of the gravitational waves: 
\begin{equation}
\left(\frac{\partial^{2}\Gamma_{\nu,\alpha}^{\lambda}}{\partial^{2}x}-\frac{\partial^{2}\Gamma_{\nu,\alpha}^{\lambda}}{c^{2}\partial^{2}t}\right)V_{\lambda}=R_{\nu\alpha}-J_{\nu\alpha}\label{eq: justifies}
\end{equation}

\section{Conclusion remarks}

We have shown that the non-conservative gravitational field is similar
to a wave for 64 variables $\Gamma_{\nu\alpha}^{\lambda}$ in a non
linear material where we have complex non linear phenomena inside
the virtual material that represents the vacuum. In previous equations,
in the free field of the medium are present the Proca terms $\Gamma_{\rho}\Gamma_{\lambda}$
the Chern-Simons terms $\left(\partial_{\nu}\Gamma_{\rho}\right)\Gamma_{\lambda}$
and the Maxwell-like terms $\left(\partial_{\nu}\Gamma_{\rho}\right)\left(\partial_{\mu}\Gamma_{\lambda}\right)$.
Hence, we have the mass terms, the topologic terms and the like electromagnetic
field terms. We can model the gravitational wave with torsion as particle
in a non linear medium which gives the mass of the particle, in a
way that can be compared to usual SSB processes of the standard model
\cite{key-34}.

Finally, we take the chance to stress that one gets a great intellectual
and aesthetic pleasure by verifying the deep internal coherence between
syntax and semantics in a physical theory. In the current approach,
we have seen that the gauge philosophy of General Relativity can include
torsion without ``ad hoc'' assumptions, differently from other theories
where torsion is forcedly inserted, see the review \cite{key-22}
for details. This issue improves our confidence on the power of symmetry
in physics. One indeed argues that, differently from specific models,
the most fundamental physical theories can be captured through some
symmetry principle. This is the same direction of the formal unification
between General Relativity and Quantum Mechanics \cite{key-23}. In
his famous Essay \cite{key-39}, Kepler prophetically attributed the
symmetries of the snowflake to its constituents. On one hand, maybe
that only a future quantum theory of gravity could explain new behaviors
of symmetry as \textquotedbl{}emergent\textquotedbl{}. On the other
hand, researchers can work having a full confidence on a strong complementarity
between \textquotedbl{}the bricks of the Universe\textquotedbl{} and
\textquotedbl{}symmetry\textquotedbl{}. This is, perhaps, the richest
and most intriguing speculative character of Einstein's legacy.

\section{Acknowledgements }

The authors thank an unknown referee for useful comments. Christian
Corda has been supported financially by the Research Institute for
Astronomy and Astrophysics of Maragha (RIAAM), Iran, project number
1/4717-111.


\begin{thebibliography}{10}
\bibitem{key-1}E. Noether, Invariante Variationsprobleme, Nachr.
v. d. Ges. d. Wiss. zu Göttingen 235 (1918). 

\bibitem{key-2}C. Quigg, Gauge Theories of the Strong, Weak, and
Electromagnetic Interactions, Princeton Univ. Press, 2nd edition,
2013. 

\bibitem{key-3}R. Healey, Gauging What's Real: The Conceptual Foundations
of Contemporary Gauge Theories, Oxford Univ. Press, 2009. 

\bibitem{key-4}R. Mignani , E. Pessa , G. Resconi, Non-Conservative
Gravitational Equation, General Relativity and Gravitation, Vol.29,
No.8, 1049-1073 (1997). 

\bibitem{key-5}R. Mignani , E. Pessa , G. Resconi, Electromagnetic-Like
Generation of Unified-Gauge Theories, Phys. Essay, 12(1) 62-79 (1999). 

\bibitem{key-6}M. Hamani Daouda, M. E. Rodrigues, M. J. S. Houndjo,
New Black Holes Solutions in a Modified Gravity, ISRN Astronomy and
Astrophysics, Article ID 341919 (2011). 

\bibitem{key-7}I. Licata, Methexis, Mimesis and Self Duality: Theoretical
Physics as Formal Systems, Versus, 118, 119-140 (2014). 

\bibitem[8]{key-8}B. de Wit, H. Nicolai, H. Samtleben, Gauged supergravities,
tensor hierarchies, and M-theory, JHEP02, 044 (2008). 

\bibitem[9]{key-9}Gutfreund, H., Renn, J. The Road to Relativity:
The History and Meaning of Einstein's \textquotedbl{}The Foundation
of General Relativity\textquotedbl{} Featuring the Original Manuscript
of Einstein\textquoteright s Masterpiece, Princeton Univ. Press (2015).

\bibitem[10]{key-10}Smolin, L. Lessons from Einstein's 1915 discovery
of general relativity, arXiv:1512.07551 {[}physics.hist-ph{]}.

\bibitem[11]{key-11}Brian Pitts, J. Universally Coupled Massive Gravity,
III: dRGT-Maheshwari Pure Spin-2, Ogievetsky-Polubarinov and Arbitrary
Mass Terms, Annals of Physics 365, 73 (2016).

\bibitem[12]{key-12}Duetsch,M. Massive vector bosons: Is the geometrical
interpretation as a spontaneously broken gauge theory possible at
all scales? Reviews in Mathematical Physics 27(10):1550024 (2016).

\bibitem[13]{key-13}Duetsch,M., Higgs Mechanism and Renormalization
Group Flow: Are They Compatible? arXiv:1502.00099 {[}hep-th{]} (2015).

\bibitem[14]{key-14}Pessa, E. The Concept of Particle in Quantum
Field Theory, in Vision of Oneness, I. Licata \& A. Sakaji Eds., Aracne,
Rome (2011), also in arXiv:0907.0178 {[}physics.gen-ph{]} (2009).

\bibitem[15]{key-15}Colosi, D., Rovelli, C.: What is a Particle?
Class. Quantum Grav. 26 025002 (2009).

\bibitem[16]{key-16}Davies, P. \textquoteleft Particles do not exist\textquoteright{}
in S. M. Christensen (Ed.), Quantum Theory of Gravity, Adam Hilger,
p. 66 (1984). 

\bibitem[17]{key-17}Manning, A.G.,Khakimov, R.I. ,Dall , R. G., Truscott,
A. G., Wheeler's delayed-choice gedanken experiment with a single
atom, Nature Physics11, 539 (2015).

\bibitem[18]{key-18}Butterfield, J. Reduction, Emergence and Renormalization,
The Journal of Philosophy, volume 111, 5 (2014).

\bibitem[19]{key-19}Meschini, D. Planck-Scale Physics: Facts and
Beliefs, Found. Of Sc. 12, 4, 277 (2007). 

\bibitem[20]{key-20}Blasone, M., Jizba, P., Scardigli, F., Vitiello,
G. Dissipation and quantization for composite systems, Phys.Lett.A373:4106
(2009). 

\bibitem[21]{key-21}Chen Nin Yang, Einstein's Impact on Theoretical
Physics. Physics Today 33, 42 (l980). 

\bibitem[22]{key-22}Capozziello, S., DeLaurentis, M. Extended Theories
of Gravity, Phys. Rep. 509, 4-5, (2011).

\bibitem[23]{key-23}Resconi, G., Licata, I., Fiscaletti, D. Unification
of Quantum and Gravity by Non Classical Information Entropy Space,
Entropy 15(9), 3602(2013). 

\bibitem[24]{key-24}Resconi, G., Licata, I., Beyond an Input/Output
Paradigm for Systems: Design Systems by Intrinsic Geometry, Systems,
2(4), 661 (2014)

\bibitem[25]{key-25}{[}25{]} Corda, C. Interferometric detection
of gravitational waves: the definitive test for general relativity,
Int. Journ. Mod. Phys. D, 18, (2009). 

\bibitem[26]{key-26}Cartan, E. Sur une généralisation de la notion
de courbure de Riemann et les espaces à torsion, C. R. Acad. Sci.
(Paris) 174, 593 (1922). 

\bibitem[27]{key-27}Cartan,E. Sur les variétés à connexion affine
et la théorie de la relativité généralisée, Part I: Ann. Éc. Norm.
40, 325 (1923) and ibid. 41, 1 (1924); Part II: ibid. 42, 17 (1925). 

\bibitem[29]{key-28}Sciama, D.W. The physical structure of general
relativity, Rev. Mod. Phys. 36, 463(1964). 

\bibitem[29]{key-29}Kibble,T.W.B. Lorentz invariance and the gravitational
field, J. Math. Phys. 2, 212 (1961). 

\bibitem[30]{key-30}Einstein, E. Riemann-Geometrie mit Aufrechterhaltung
des Begriffes des Fernparallelismus, Preussische Akademie der Wissenschaften,
Phys.-math. Klasse, Sitzungsberichte 217 ( 1928). 

\bibitem[31]{key-31}Pop\l awski, N. Nonsingular, big-bounce cosmology
from spinor-torsion coupling, Phys. Rev. D 85, 107502 (2012).

\bibitem[32]{key-32}Hammond, R. T. The necessity of torsion in gravity,
Int. Journ. Mod. Phys. D, 19, 2413 (2010). 

\bibitem[33]{key-33}Kleinert, H. The Invisibility of Torsion in Gravity,
users.physik.fu-berlin.de/\textasciitilde{}kleinert/386/386.pdf. 

\bibitem[34]{key-34}Mukhi, S. Unravelling the novel Higgs mechanism
in (2+1)d Chern-Simons theories, Jour. of High Energy Phys., 83 (2011). 

\bibitem[35]{key-35}Misner, C. W., Thorne, K. S. and Wheeler, J.
A. Gravitation\emph{,} W. H. Feeman and Company (1973).

\bibitem[36]{key-36}Landau, L.D. \& Lifshitz, E.M. The Classical
Theory of Fields (Volume 2 of A Course of Theoretical Physics) Pergamon
Press 1971.

\bibitem[37]{key-37}Lorenz, L. Phil. Mag. 34, 287 (1867). 

\bibitem[38]{key-38}Mignani R., Pessa E., Resconi G., Nuovo Cimento
B, 108, 1319 (1993).

\bibitem[39]{key-39}J. Kepler, De nive sexangula (On the Six-Cornered
Snowflake, 1611).
\end{thebibliography}
\end{document}